\documentclass[final,3p, article,times,10pt]{elsarticle}
\pdfoutput=1

\usepackage{amsthm}
\usepackage{lineno,hyperref}
\modulolinenumbers[1]
\usepackage{natbib}
\journal{arxiv.org}

\usepackage{amssymb}

\usepackage{amsmath}
\usepackage{gensymb}
\usepackage[]{units}
\usepackage{tabularx}
\usepackage{tabulary}
\usepackage{float}
\usepackage{bibentry}
\usepackage{graphicx}

\begin{document}

\begin{frontmatter}


\title{Characterisation of carbon fibre-reinforced polymer composites through complex Radon-transform analysis of eddy-current data \tnoteref{label1}}
\tnotetext[label1]{}
\author{R.R. Hughes\corref{cor1}\fnref{label2}}
\ead{robert.hughes@bristol.ac.uk}

\title{Characterisation of carbon fibre-reinforced polymer composites through complex Radon-transform analysis of eddy-current data}

\author{B.W. Drinkwater$^1$ and R.A. Smith$^1$}
\address{$^1$Department of Mechanical Engineering, Queens Building, University Walk, University of Bristol, BS8 1TR, UK}

\begin{abstract}
Maintaining the correct fibre orientations and stacking sequence in carbon-fibre reinforced polymers (CFRP) during manufacture is essential for achieving the required mechanical properties of a component. This paper presents and evaluates a method for the rapid characterisation of the fibre orientations present in CFRP structures, and the differentiation of different stacking sequences, through the Radon-transform analysis of complex-valued eddy-current testing (ECT) inspection data. A high-frequency (20 MHz) eddy-current inspection system was used to obtain 2D scans of a range of CFRP samples of differing ply stacking sequences. The complex electrical impedance scan data was analysed using Radon-transform techniques to quickly and simply determine the dominant fibre orientations present in the structure. This method is compared to 2D-fast Fourier transform (2D-FFT) analysis of the same data and shown to give superior quantitative results with comparatively fewer computational steps and corrections. Further analysis is presented demonstrating and examining a method for preserving the complex information inherent within the eddy-current scan data during Radon-transform analysis. This investigation shows that the real and imaginary components of the ECT data encode information about the sacking sequence allowing the distinction between composites with different stacking structures. This new analysis technique could be used for in-process analysis of CFRP structures as a more accurate characterisation method, reducing the chance of costly manufacturing errors. 
\end{abstract}


\end{frontmatter}


\section{Introduction}\label{sec:intro}
	Carbon-fibre reinforced polymer (CFRP) is used in increasingly sophisticated industrial applications \citep{Roberts2007}. High strength-to-weight ratio, tolerance to fatigue damage and the ability to form complex geometries make it an attractive material for use in many industries including aerospace and automotive. CFRP components get their strength from the order and alignment of fibres along specific axes \cite{Mallick2007} and as such their structural integrity is dependent on reliable manufacturing processes. Misalignments and stacking errors can easily occur during the layup process of complex geometries and may lead to in-plane fibre waviness, out-of-plane ply wrinkling and ply-bridging in the finished component, introducing structural weaknesses \cite{Mallick2007, Yokozeki2006, Kawai2004, Hancox1975}. Costly manufacturing errors such as these are easily corrected if found during the layup process. As such, new reliable non-destructive testing (NDT) techniques are required for quality-control on these parts during the layup process.
	
	The introduction of foreign contaminants in the manufacturing process must be minimised to prevent defects in the finished component. This presents a problem for many traditionally effective NDT methods such as ultrasonic testing (UT) as they require contact with the test material through a coupling medium or for the sample to be submerged in water. Potential in-line non-destructive testing techniques include optical sensing with image processing methods to measure fibre alignment \cite{Vanclooster2009, Tunak2014, Lomov2008, Zambal2015}. Such approaches are fast and low cost but cannot measure orientation in sub-surface plies. However, an alternative lies in eddy-current testing (ECT), a non-contact inductive sensing NDT technique that can penetrate below the surface. 
	
	ECT measures the local electrical properties of a sample's surface and near surface. The technique works by measuring the complex electrical impedance ($Z = R + iX$) of an electromagnetic coil when excited with a sinusoidal signal. $R$ is the real, or resistive, component and $X$ is the imaginary, or reactive, component of impedance. The measured impedance depends on the induced current density, $J(z)$, within the material which decays with depth, $z$, into an isotropic material as \citep{Wheeler1942},

\begin{equation}
J = J_0 exp\left(-\frac{z}{\delta_{sd}} (1+i)\right).
\end{equation}\label{eqn:J}
	 
	 The depth at which the magnitude of the current density has fallen to 1/e f its surface value is know as the depth of penetration, $\delta_{sd}$, and is dependent on the electrical conductivity, $\sigma$ and magnetic permeability, $\mu$, of the test sample, as well as the excitation frequency, $\omega$, and the geometry of the excitation coil which is a distribution of spatial frequencies represented by the 'effective' spatial-frequency term, $\kappa$, as given below \citep{Dodd1968},
	
	\begin{equation}
\delta_{sd} = \frac{1}{\sqrt{\kappa^2 + \mu\sigma\omega}}.
\end{equation}\label{eqn:skin depth}
	
	Smaller ECT coils and higher frequencies will therefore result in a faster rate of decay in a given material. In this way the depth over which an ECT measurement is made can be controlled. Although traditionally carried out on high-conductivity metallic materials, ECT techniques have been shown to be capable of detecting multiple stacked ply orientations in CFRP structures due to the electrical anisotropy of carbon-fibre plies \cite{Prakash1976, Goeje1992, Lange1994}.
	
	\subsection{Ply Orientation}
	\citet{Lange1994} recognised that a non-axially symmetric ECT probe could be rotated and used to determine ply orientations in a sample by deconvolving the angular measurements via a Fourier-transform. This was later expanded for use on 2D raster-scan ECT data by performing 2D fast Fourier transforms (2D-FFTs)  \citep{Mook2001}. This analysis approach has been widely adopted for automated in-process monitoring of fibre alignment \citep{Bardl2016, Heuer2013, Ayres2008, Cheng2017, Menana2009}. However, 2D-FFT analysis is an inherently Cartesian transform making the accurate extraction of \textit{angular} information an inelegant secondary process requiring careful correction to remain universally valid.
	
\begin{figure}[h]
\centering
    \includegraphics[width=0.9\textwidth]{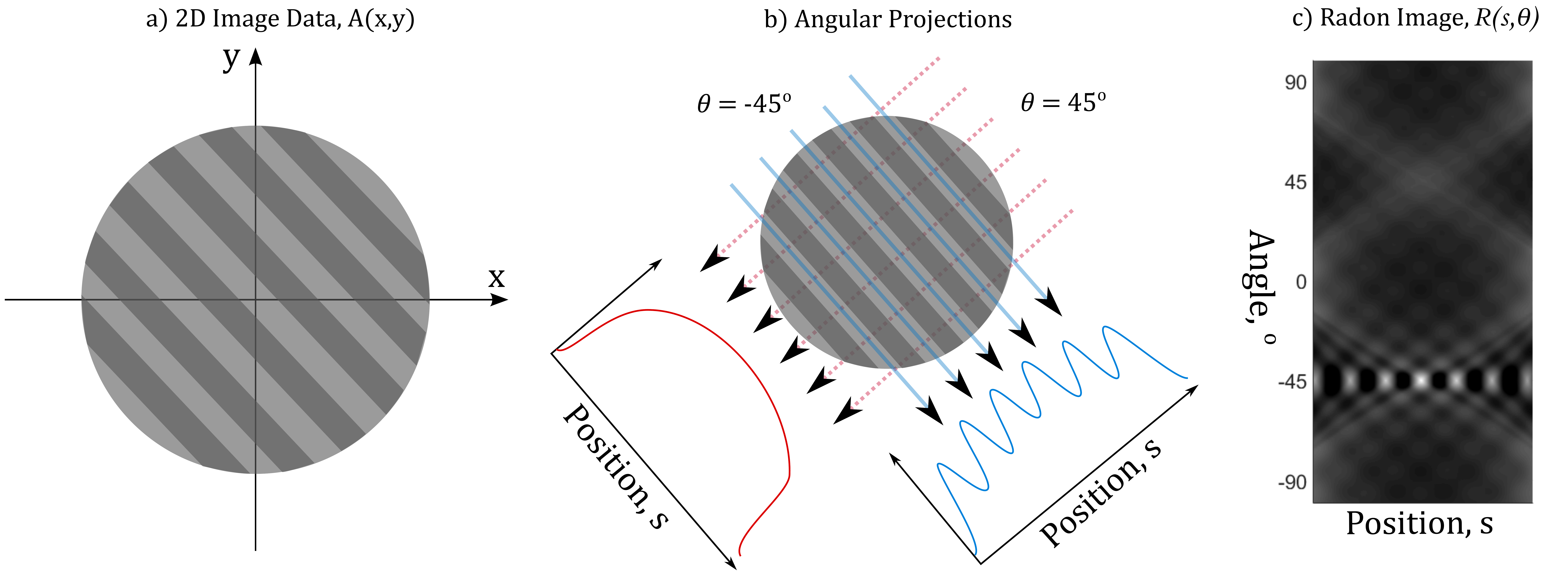}\caption{Schematic diagram showing the principle of Radon transform analysis of a structured 2D image dataset (a), via pixel summation along different projection angles. c) shows the full angular radon image, demonstrating image structure at $-45\degree$.}
    \label{fig:Radon Theory}
\end{figure}

	A potential alternative to the 2D-FFT method is Radon-transform (RT) analysis, which has been used in a wide range of image processing applications for extracting or reconstructing angular information \citep{Schaub2013, Seo2004, Jafari-Khouzani2005, Krause2010, Zhang2007, Smith2009}. A Radon transform performs line integrals in the plane of a 2D image function, $A(x,y)$, mapping line-integrals at position $s$, and angle $\theta$, into $(s,\theta)$-space (Figure~\ref{fig:Radon Theory}). The resulting transform image, $R(s,\theta)$, contains information on the angular structure of an image function $A(x,y)$. The Radon transform image can be straightforwardly analysed to determine which angles exhibit strong structural information with fewer processing steps compared to 2D-FFT techniques \citep{Smith2015}.
	
	\subsection{Ply Structure}
	A desirable output from quality-control inspections of CFRP structures is the stacking sequence. Previous authors using ECT typically perform the standard 2D-FFT analysis on either the magnitude, $|Z|$, or the phase, $\psi$, of the ECT data,
	
	\begin{equation}\label{magNpha}
\begin{split}
|Z| & = \sqrt{R^2 + X^2} \\
\psi & = \tan^{-1} \left( \frac{X}{R} \right),
\end{split}
\end{equation}

	 and use the signal strength at particular orientations to infer stacking sequences making the assumption that weaker orientation signals are deeper in the material \citep{Schmidt2014, Bardl2016}.  This approach overlooks the full information encoded into the complex values and could misrepresent the structure of more complex stacking sequences. 
	 
	 Eddy-current sensors measure the superposition of magnetic field contributions from all depths, arising from the flow of current within the material. The complex current density, $J(z)$, decays in magnitude and phase with depth such that each layer of a material will contribute differently to the real and imaginary parts of the overall magnetic-field measured and thus the ECT coil impedance.	The electrical anisotropy of CFRP plies leads to a non-uniform distribution of current density within cross-ply CFRP materials, as shown by \citet{Cheng2017} through finite-element modelling. The simulated results in \citet{Cheng2017} predict that the current density is a local maximum at the interface between cross-ply layers, where contact is made between orthogonal fibres. It is predicted that the complex phase of the current density will also vary irregularly with depth and exhibit transitions at the ply interfaces. The difference in behaviour of the current density in different CFRP stacking sequences will therefore lead to measurable differences in the complex contributions measured by the ECT sensor.
\\

	In this study, RT analysis is evaluated as an alternative to 2D-FFT analysis for the non-destructive characterisation of fibre orientations in CFRP structures. This RT method is then used to carry out the analysis on the full complex ECT data from various different samples to explore the effect of stacking sequence of ECT measurements and exploit these effects to distinguish between different CFRP structures.
%
%

\section{Materials and Methods}
CFRP samples, with different stacking sequences, were manufactured and tested using a high-frequency eddy-current inspection system.  The complex coil-impedance data was analysed using 2D-FFT \& RT analysis to explore the effects of ply orientations and lay-up structure.

	\subsection{Test Specimens}\label{sec:test bits}
	Three IM7/8552 CFRP specimens were manufactured in-house with specific ply structures - unidirectional, cross-ply and complex - shown in Table~\ref{tab:ply lay-up}.
	
\begin{table}[H]
	\caption[]{Test specimen ply lay-up structures}
	\begin{center}\small
		\begin{tabular*}{0.5\textwidth}{@{\extracolsep{\fill} } l c c}
			\hline
			\hline
			Sample & No. Plies & lay-up structure \\
			\hline
			001 & 32 & [$90^o$]$_{32}$ \\
			002 & 32 & [$0^o$/$90^o$]$_{16}$ \\
			003 & 32 & [$90^o$/$45^o$/$90^o$/$135^o$]$_{8}$  \\
			\hline
		\end{tabular*}
		\label{tab:ply lay-up}
	\end{center}
\end{table}
Samples were manufactured manually from pre-preg plies and cured, as per the manufacturer's instructions \citep{Hexcel2016}, under vacuum and sandwiched between aluminium plates. The resulting samples were 4 mm thick.
%

	\subsection{Experimental Setup} \label{sec:setup}
\begin{figure}[h]
\centering
    \includegraphics[width=0.5\textwidth]{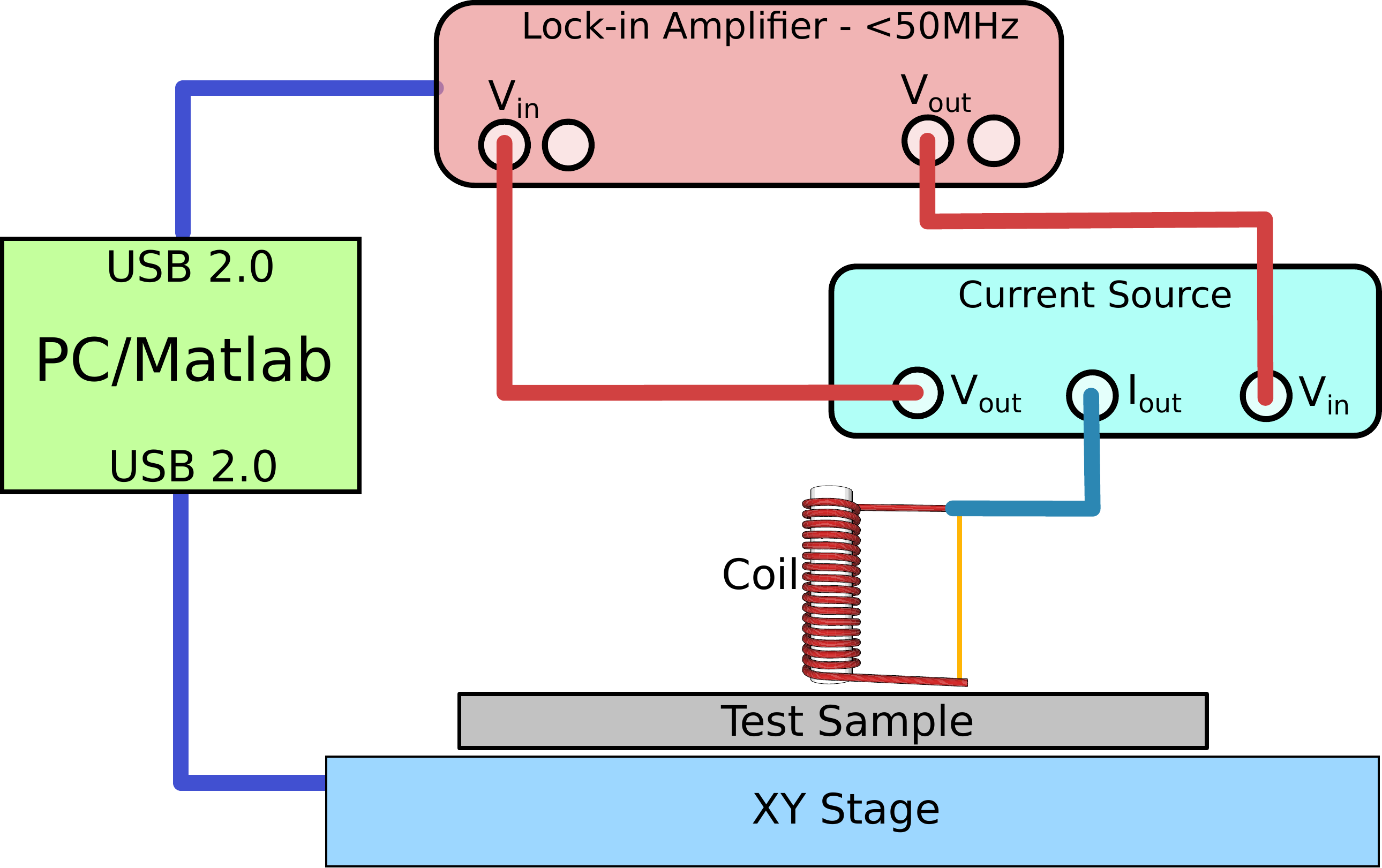}
    \caption{Schematic diagram of the experimental set-up for eddy-current measurements.}
    \label{fig:Expmnt setup}
\end{figure}

	A Zurich Instruments HFLI2 Lock-in Amplifier was used to operate and measure the electrical impedance of the ECT coil at 20 MHz. A Howland current source\footnote{Designed and manufactured by SonEMAT (Coventry, UK)} was used to convert the input voltage into an equivalent current to the sensor coil \citep{Hughes2014}, and mounted immediately behind the sensor coil to eliminate the detrimental high-frequency impedance effects of connecting cables. A low inductance, 25 turn, $1$ mm external diameter coil was constructed in-house around a $0.75$ mm diameter ferrite rod core\footnote{Grade 61 from Fair-Rite (New-York, US)}. A schematic diagram of the experimental set-up is shown in Figure~\ref{fig:Expmnt setup}.
	
	Two-dimensional raster scans were performed on test samples using two high-accuracy linear stages controlled via PC. Scan data consisted of complex coil-impedance measurements taken at X and Y positions in $0.2$ mm increments over a $20\times20$ mm area. The ECT probe was used in an absolute (reflection) mode \citep{BlitzEM} and Teflon tape used to protect the probe head during scanning, providing a constant lift-off  of $0.1$ mm from the surface, maintained by the pressure of a spring-loaded probe mount. Initial investigations revealed that single-frequency excitation at 20 MHz produced the highest image contrast for the test samples and so was used throughout the study although the frequency dependence of the response will be investigated in future studies.
	
\subsection{ECT Data}
\begin{figure}[h]
\centering
    \includegraphics[width=0.75\textwidth]{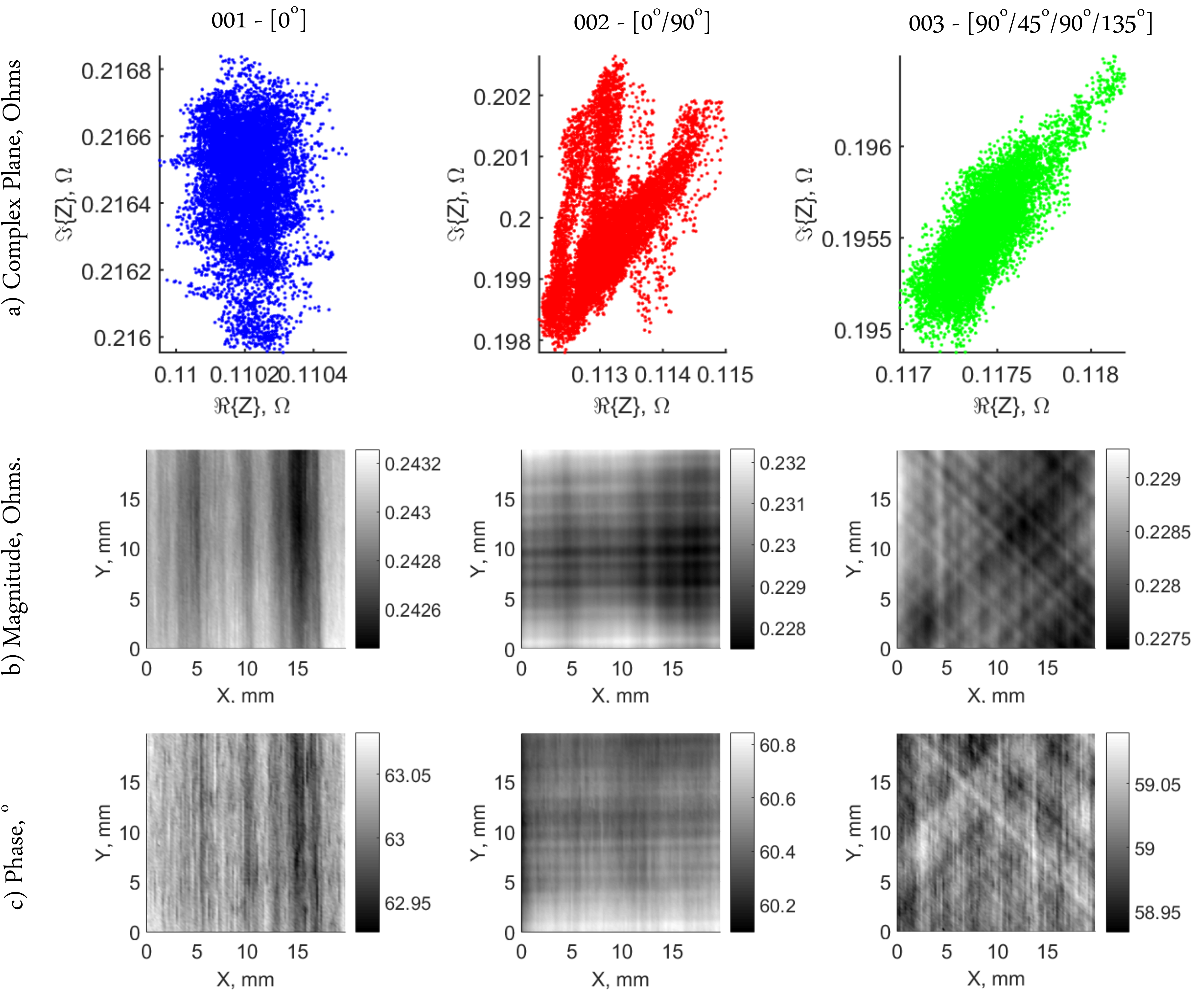}\caption{Raw 20 MHz ECT scan data of test samples (left to right) 001, 002, 003 showing the data in the complex plane (a), and in 2D images of the magnitude (b) and phase (c).}
    \label{fig:Raw C-scans 1}
\end{figure}

The eddy-current scans produce 2D $M\times N$ complex matrices, $\tilde{A}(x,y)$. Prior to analysing the fibre orientations, a 41$\times$41 pixel ($8.2\times 8.2$ mm) hamming-window high-pass filter was applied to the scan data to remove low spatial-frequency variations. The unfiltered complex data is shown in the complex plane in Figure~\ref{fig:Raw C-scans 1}.a, alongside magnitude (b) and phase (c) images.
		
Figure~\ref{fig:Raw C-scans 1}.a shows the complex-plane impedance ($Z$) data for the full 2D data-sets for each sample. We hypothesis that the data distribution in complex-space provides information about the structure of each material. The images shown in Figure~\ref{fig:Raw C-scans 1}.b \& c clearly show the fibre orientations within the three test samples but not the depth of plies. Reliable automated analysis is therefore required to quantify the orientations present, their dominance in the data and, if possible, their stacking sequence.

\section{Ply Orientation}
\begin{figure}[h]
\centering
    \includegraphics[width=0.75\textwidth]{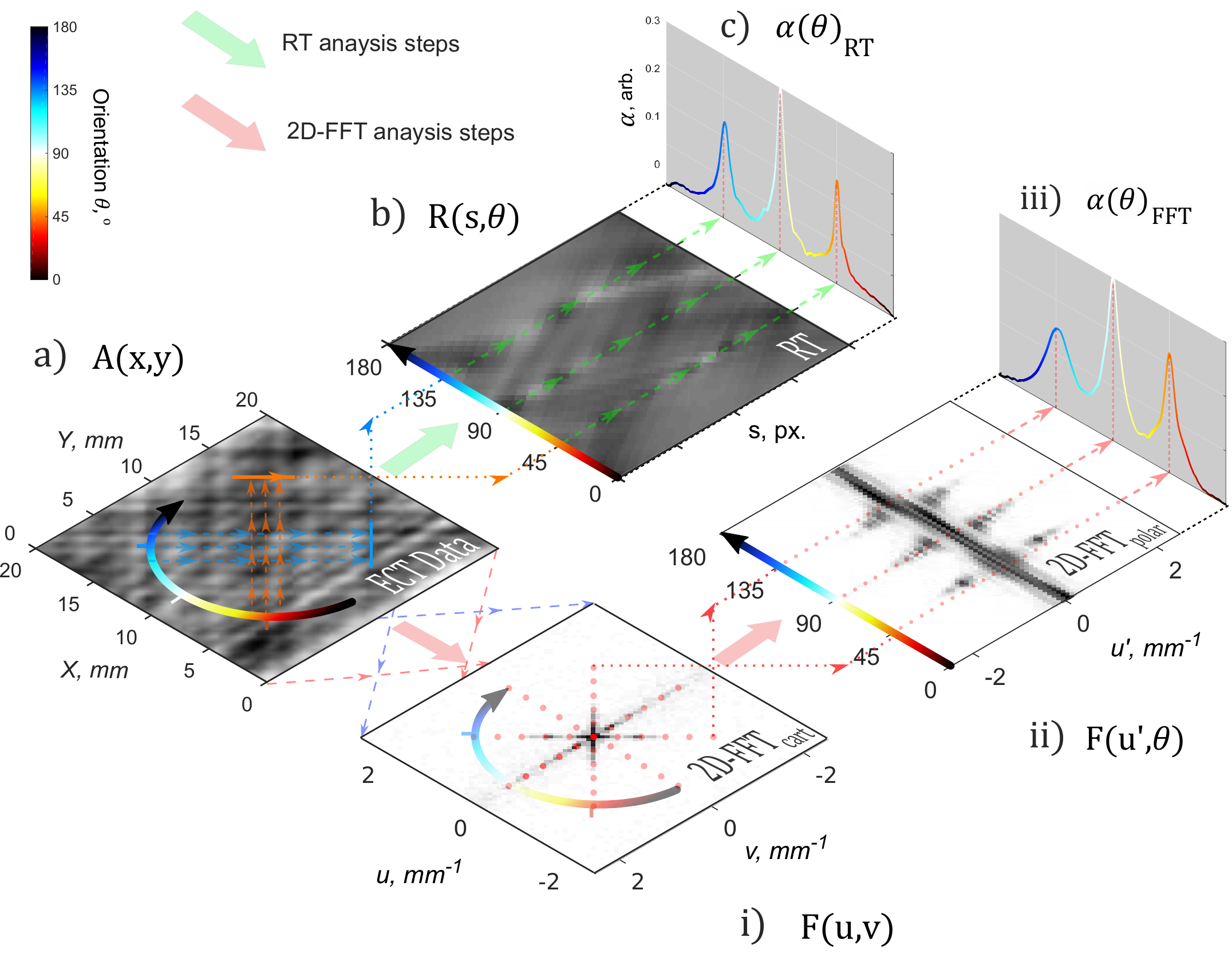}\caption{Comparison between Radon transform (RT) and 2D-FFT analysis steps of ECT data (a) for quantitative characterisation of ply orientations. The RT process (a-c) involves performing an RT to give the RT image (b) followed by the calculation of the angular-distribution (c). The 2D-FFT process (a, i-iii) involves transforming (a) into its 2D-FFT image (i), re-sampling in polar co-ordinates (ii) and generation of angular-distribution (iii).}
    \label{fig:RadonVisual}
\end{figure}

Radon-transform analysis was evaluated against the current literature standard, 2D-FFT, as a method for automatically detecting and quantifying the dominant ply orientations present in test samples. For this comparison, the magnitude of the complex ECT was used for both analysis techniques and angular-distribution plots obtained from each processing method. Figure~\ref{fig:RadonVisual} shows the steps involved in the RT (a-c) and 2D-FFT (a \& i-iii) analysis processes.

\subsection{2D-FFT Analysis}
	2D-FFT analysis was performed on the ECT magnitude scan data, $|\tilde{A}(x,y)|$, (Figure~\ref{fig:RadonVisual}.a) of each test sample (see Table~\ref{tab:ply lay-up}) to give the 2D-FFT image, $F(u,v)$, (Figure~\ref{fig:RadonVisual}.i). The 2D-FFT of an $M\times N$ image $A(x,y)$ is,
	\begin{equation}
	F(u,v) = \frac{1}{MN} \sum^{M} _{x=0} \sum^{N} _{y=0} A(x,y) \exp\left[-i 2\pi\left(\frac{ux}{M}+ \frac{vy}{N}\right)\right].
	\label{eqn:2DFFT}
	\end{equation}
	
The center pixel is removed from the $F(u,v)$ image and the image re-sampled in polar coordinates using cubic interpolation at radial increments of 0.2 mm and angular increments of $0.1^{\circ}$ to give the polar 2D-FFT, $F(u',\theta)$, (Figure~\ref{fig:RadonVisual}.ii) \citep{Averbuch2006}. Angular distributions, $\alpha(\theta)$, were calculated by summing along the radial axis, $u'$, for each angle, $\theta$, as performed by \citet{Bardl2016}. The 2D-FFT images are shown in Figure~\ref{fig:2DFFT ims}.b and the resulting angular-distributions, $\alpha{(\theta)}_{FFT}$, are shown in Figure~\ref{fig:2DFFT ims}.d, showing peaks for the dominant fibre orientations within each sample. 

\subsection{Radon Transform Analysis} \label{sec:Radon theory}
Radon transform analysis was performed on the ECT magnitude scan data (Figure~\ref{fig:RadonVisual}.a-c) and compared against the 2D-FFT analysis method. The Radon transform is given as,

\begin{equation}
R(s,\theta) = \sum^{M-1} _{x=0} \sum^{N-1} _{y=0} A(x,y) \delta(x \cos \theta + y \sin \theta - s).
\label{eqn:Radon}
\end{equation}

In equation~\ref{eqn:Radon}, $\delta$, is a Dirac delta function \citep{Schaub2013} allowing for the summation of values along the orientation, $\theta$, direction. The angular distribution, $\alpha(\theta)_{RT}$, was calculated from the RT image (Figure~\ref{fig:RadonVisual}.b) by the mean absolute gradient along the $s$-axis at each orientation angle, $\theta$.

\begin{equation}
\alpha(\theta)_{RT} = \frac{1}{N} \sum^{N} _{n=1}\ {\left\vert \frac{\partial {R}} {\partial s}\right\vert_n},
\label{eqn:ang dist}
\end{equation}
	
where $N$ is the number of points in the position axis $s$. The resulting angular distribution, $\alpha(\theta)$, shows the orientation and dominance of certain ply orientations in the original image data, $A(x,y)$, (see Figure~\ref{fig:RadonVisual}.c). 

\subsection{Technique Comparison} \label{sec:compare}
\begin{figure}[h]
\centering
    \includegraphics[width=1.0\textwidth]{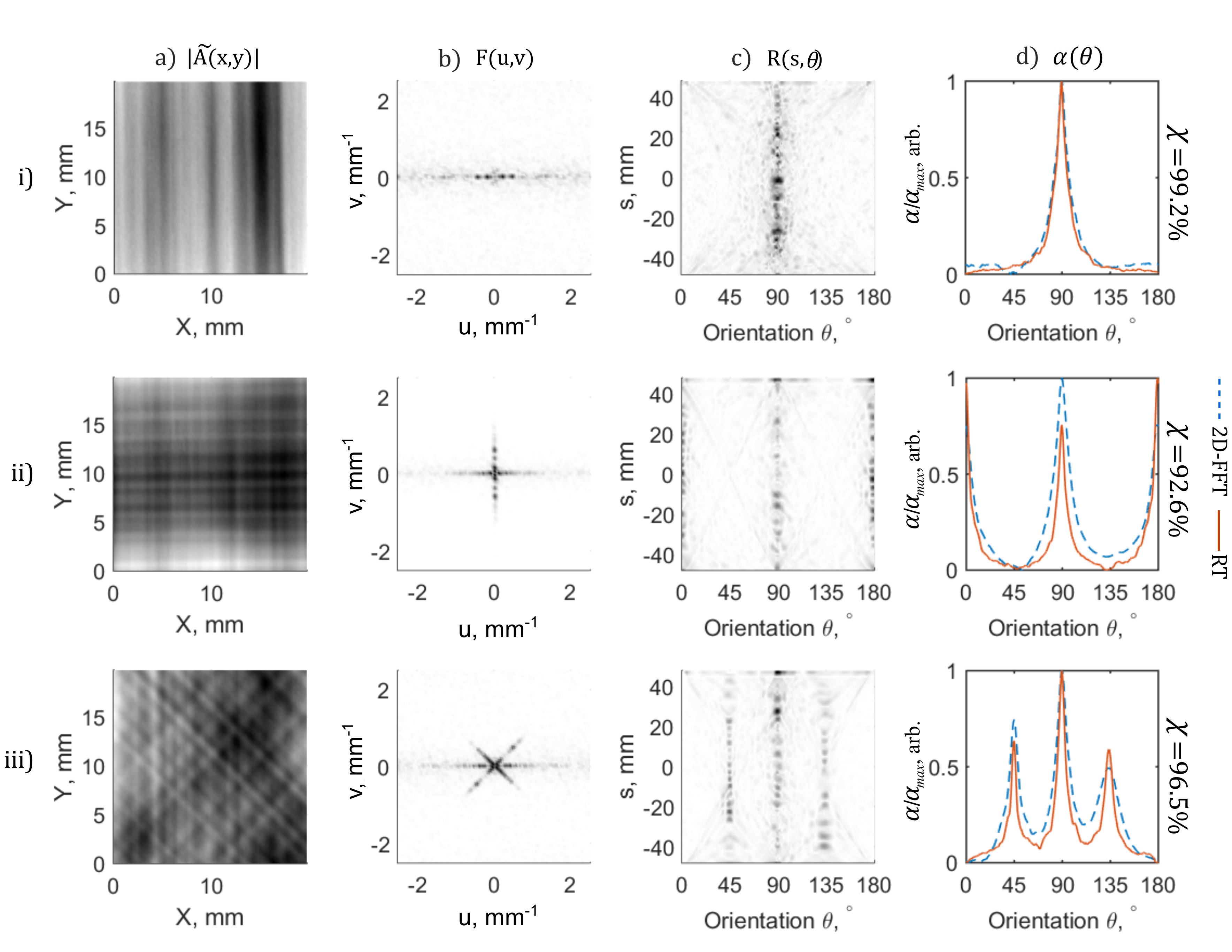}\caption{Image-processing results of ECT scan data showing a) absolute magnitude, $|\tilde{A}(x,y)|$, of the scan data, b) the 2D Fourier transform (2D-FFT) image, $F(u,v)$, c) the Radon transform (RT) image, $R(s,\theta)$, and d) the normalised angular-distributions obtained from 2D-FFT and RT images. Cross-correlation coefficients, $\chi$, between the two angular distributions are displayed next to each angular-distribution plot. Results for three test samples are shown i) 001 - [$90^{\circ}$], ii) 002 - [$0^{\circ}/90^{\circ}$] and iii) 003 - [$90^{\circ}/45^{\circ}/90^{\circ}/135^{\circ}$].}
    \label{fig:2DFFT ims}
\end{figure}

Figure~\ref{fig:2DFFT ims} directly compares the results of the two analysis methods and plots their angular distributions in the same plot, normalised to their peak magnitude $\alpha_{max}$. In this way the angular distributions are normalised to the dominant (surface) ply orientation. The features of peaks at specific angles provide information as to the structure and alignment reliability of the lay-up process. The peak features for each analysis technique are compared in Table~\ref{tab:compare}.

\begin{table}[h]
\caption[]{Angular distribution peak features. Comparison between 2D-FFT and Radon transform (RT) analysis results.}
\begin{center}\small	
\begin{tabular}{*8c} 
			{} & {} & \multicolumn{2}{c} {\textbf{Orientation, $\theta \pm 0.1^{\circ}$ }} & \multicolumn{2}{c}{\textbf{Relative Magnitude} } & \multicolumn{2}{c}{\textbf{FWHM $\pm 0.2^{\circ}$}} \\
			\hline
			Sample & {} &  2D-FFT & RT & 2D-FFT & RT & 2D-FFT & RT \\
			\hline
			\hline
			001 & [90] & 89.7 & 89.6 & 1.00 & 1.00 & 17.5 & 11.6 \\
			\hline
			002 & [0] & -0.7 & -0.6 & 0.76 & 1.00 & 26.3 & 7.6 \\
			{} & [90] & 89.5 & 89.5 & 1.00 & 0.76 & 21.9 & 8.7 \\
			\hline
			003 & [45] & 44.8 & 44.7 & 0.74 & 0.63 & 25.7 & 6.1 \\
			{} & [90] & 89.6 & 89.6 & 1.00 & 1.00 & 20.9 & 8.4 \\
			{} & [135] & 133.3 & 133.7 & 0.49 & 0.59 & 48.2 & 10.5 \\
			\hline
		\end{tabular}
		\label{tab:compare}
	\end{center}
\end{table}

The results shown in Figure~\ref{fig:2DFFT ims} and Table~\ref{tab:compare} compare the RT analysis method to 2D-FFT analysis. From Figure~\ref{fig:2DFFT ims} there is good general qualitative agreement between both techniques in measuring the dominant orientations present in the ECT data. Cross-correlation coefficients, $\chi$, were calculated between the normalised 2D-FFT and RT angular-distributions and displayed in Figure~\ref{fig:2DFFT ims}.d to quantify the agreement between the techniques. The major differences between the two analysis techniques are in the relative amplitudes and angular spread of peaks in $\alpha(\theta)$. 

The angular peak spread, characterised by the full-width-half-maximum (FWHM) in Table~\ref{tab:compare}, is a measure of the alignment accuracy of different ply layers with the same orientations within the samples. Figure~\ref{fig:2DFFT ims}.d shows that RT analysis generates sharper orientation peaks compared to 2D-FFT. This is likely a result of the polar re-sampling of the 2D-FFT image (see Figure~\ref{fig:RadonVisual}.i) where the low-frequency peaks near the origin contribute to a wide range of orientations, resulting in a greater peak spread. This issue, which could obscure smaller peaks resulting from ply misalignments, is not observed in the RT analysis method. The FWHM will scale inversely with the number of pixels in the analysed data as demonstrated by \citet{Bardl2016}, tending towards a minimum limit that will depend on the diameter of the inspection coil.

\subsection{Stacking-Sequence Differentiation: Magnitude}\label{sec:AB}
\begin{figure}[h]
\centering
    \includegraphics[width=1.0\textwidth]{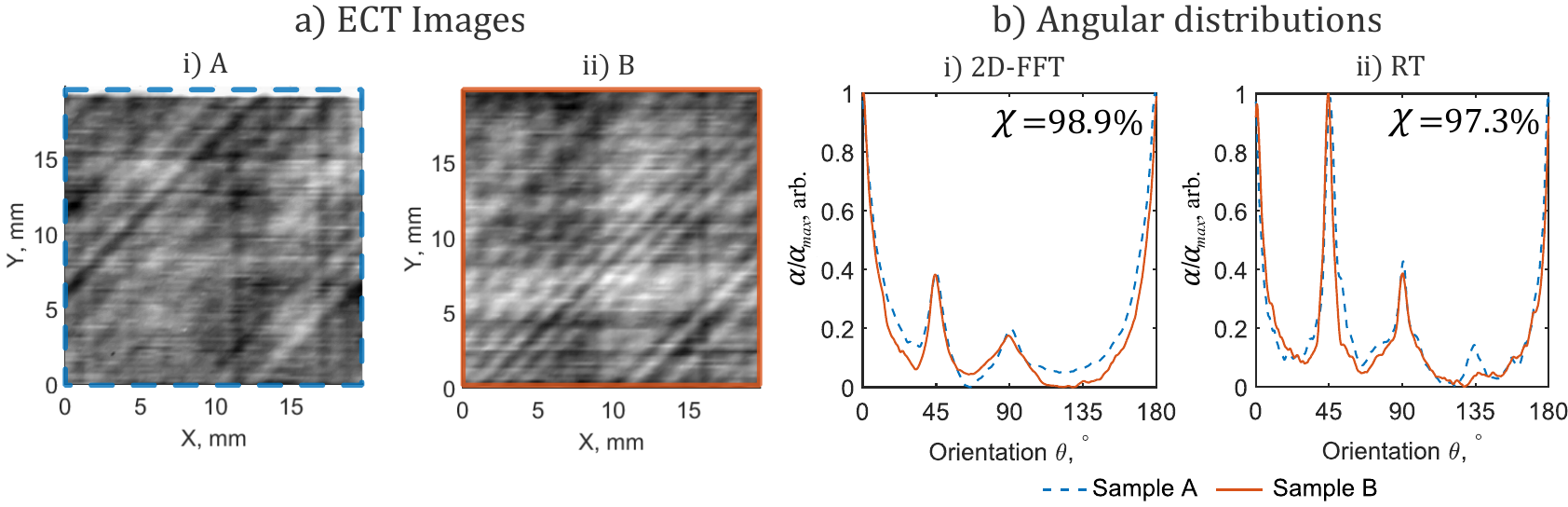}\caption{Distinguishing between similar layup structures - Comparing 2D-FFT and RT magnitude analysis of sample A - [$45^{\circ}$/$90^{\circ}$/$135^{\circ}$/$0^{\circ}$]$_{repeated}$ (blue dotted) and sample B - [$0^{\circ}$/$45^{\circ}$/$90^{\circ}$/$135^{\circ}$/$0^{\circ}$]$_{reflected}$ (red solid). Showing a) ECT magnitude images and b) normalised angular-distributions, $\alpha(\theta)$, with cross-correlation coefficients, $\chi$, between angular-distributions of the two samples from i) 2D-FFT and ii) RT analysis.}
    \label{fig:AngDistSamps}
\end{figure}

In this section two new test samples, manufactured using the same material and process as detailed in section~\ref{sec:test bits}, were inspected. Both samples contain the same number of plies but with different stacking-sequences. Samples A and B have stacking sequences [$45^{\circ}$/$90^{\circ}$/$135^{\circ}$/$0^{\circ}$]$_{repeated}$ and [$0^{\circ}$/$45^{\circ}$/$90^{\circ}$/$135^{\circ}$/$0^{\circ}$]$_{reflected}$ respectively. The specimens were inspected at 19.5 MHz using the experimental set-up in section~\ref{sec:setup} and 2D-FFT and RT angular distribution analysis performed on the magnitude of the complex ECT data. Figure~\ref{fig:AngDistSamps}.a) shows the ECT magnitude images and b) compares the normalised angular-distributions obtained via 2D-FFT and RT.

2D-FFT analysis (Figure~\ref{fig:AngDistSamps}.b.i) fails to detect the presence of plies at $135^{\circ}$ in both samples whereas RT analysis (Figure~\ref{fig:AngDistSamps}.b.ii) picks up all ply-orientations. RT analysis also produces strong $45^{\circ}$ ply peaks for both samples. This could be accounted for in sample A due to the top ply having an orientation of $45^{\circ}$. However, the reason behind the same $45^{\circ}$ strength in sample B is unclear. Cross-correlation coefficients, $\chi$, were obtained to compare the angular-distributions between samples A \& B, and show strong correlation (less than 3\% difference) between the two samples for both 2D-FFT and RT analysis methods. It is therefore reasonable to conclude that it very difficult to automatically differentiate between these two samples based on analysis of ECT magnitude data alone. 

%

\section{Complex-Component Angular Distributions}\label{sec:cmplx rad}
A truly universal quantitative evaluation system must be capable of independently validating the stacking sequence without prior knowledge. The following section explores how RT analysis can be used to better differentiate between two similar test samples via the analysis of complex information inherent in ECT measurements. As discussed in section~\ref{sec:intro}, ECT measurements produce complex values with depth information encoded into the relationship between real and imaginary components. The following study exploits this inherent property of ECT measurements to produce an angular distribution that reflects the relative changes in the complex data with depth.

\begin{figure}[h]
\centering
    \includegraphics[width=0.5\textwidth]{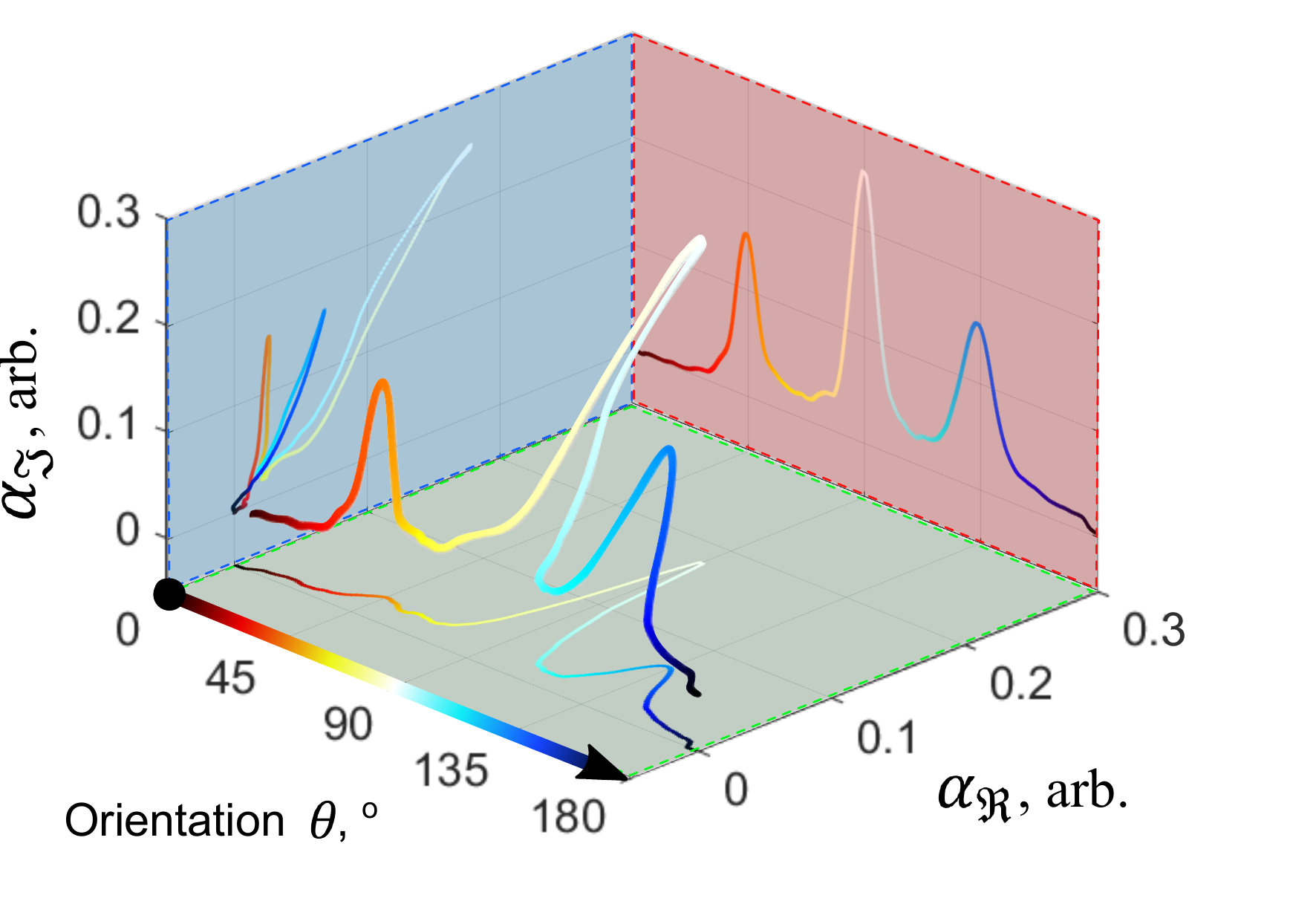}
    \caption{Example three-dimensional representation of a complex angular distribution of sample 003, obtained via Radon transform analysis. Showing the relative Real (green), Imaginary (red) and relative contribution (blue) plane projections.}
    \label{fig:demo cmplx projections}
\end{figure}

Equation~\ref{eqn:Radon}, in section~\ref{sec:Radon theory}, was used to produce the complex Radon image, $\tilde{R}(s,\theta)$, (see Figure~\ref{fig:RadonVisual}.b). The Radon image is proportional to the mean complex value along each line at orientation angle, $\theta$. In order to preserve the complex information the angular distribution calculation (equation~\ref{eqn:ang dist}) is performed on both real and imaginary components of $\tilde{R}$ separately,

	\begin{equation}
	\begin{aligned}
	{\alpha}_{\Re}(\theta) & = \frac{1}{N} \sum^{N} _{n=1}\ { \left\vert \frac{\partial {\Re\{\tilde{R}\}}} {\partial s}\right\vert_n },\\
	{\alpha}_{\Im}(\theta) & = \frac{1}{N} \sum^{N} _{n=1}\ { \left\vert \frac{\partial {\Im\{\tilde{R}\}}} {\partial s}\right\vert_n },
	\label{eqn:ang dist cmplx}
	\end{aligned}
	\end{equation}
	
where $N$ is the number of points along the position axis $s$. The resulting angular distributions, ${\alpha}_{\Re}(\theta)$ \& ${\alpha}_{\Im}(\theta)$, reflect the relative variations in the real and imaginary components of the raw complex data as a result of the material structure (i.e. fibres) at a given orientation angle, $\theta$. Figure~\ref{fig:demo cmplx projections} plots  ${\alpha}_{\Re}(\theta)$ \& ${\alpha}_{\Im}(\theta)$ relative to one another in a 3D plot.
  
\begin{figure}
\centering
    \includegraphics[width=0.75\textwidth]{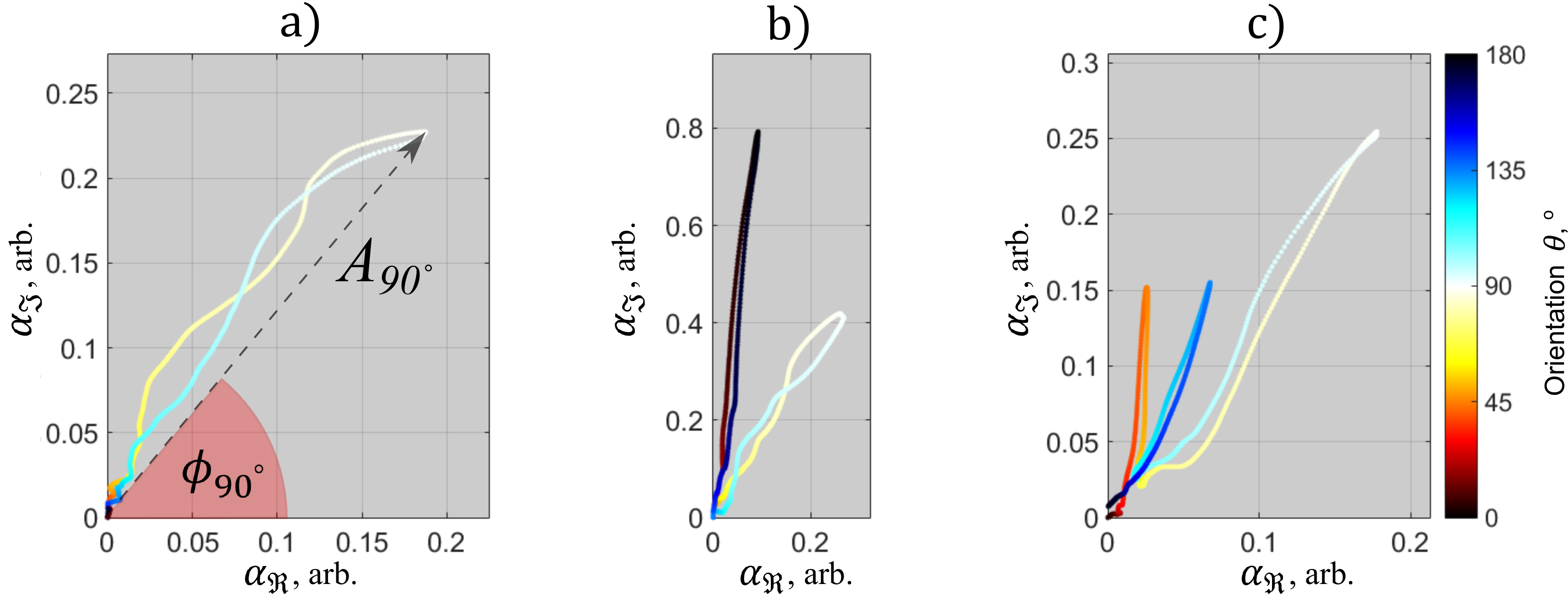}\caption{Experimental real-imaginary angular distributions of three test samples a) [$90^o$] b) [$0^o/90^o$] and c) [$90^o/45^o/90^o/135^o$] from Radon transform analysis of 20 MHz ECT inspection. The color-bar represents the orientation angle, $\theta$, in degrees,  $\phi_{\theta}$ is the phase angle and $A_{\theta}$ is the relative magnitude.}
    \label{fig:cmplx plots}
\end{figure}

The relative contribution angle (RCA), $\phi_{\theta}$, is defined as the angle a peak makes with the real axis, ${\alpha}_{\Re}(\theta)$, in the relative contribution plane (blue projection in Figure~\ref{fig:demo cmplx projections}) as demonstrated in Figure~\ref{fig:cmplx plots}.a. $\phi_{\theta}$ provides complementary information to the relative contribution magnitude (RCM), $A_{\theta}$, as to the stacking sequence of the CFRP materials.

\begin{equation}
\begin{split}
A_{\theta} & = \sqrt{\{{\alpha}_{\Re}\}^2 + \{{\alpha}_{\Im}\}^2} \\
\phi_{\theta} & = \tan^{-1} \left( \frac{{\alpha}_{\Im}}{{\alpha}_{\Re}} \right)
\end{split}
\end{equation}

The resulting relative real-imaginary plots in Figure~\ref{fig:cmplx plots} demonstrate the difference in $\phi_{\theta}$ between orientation peaks in the complex-contribution space for samples 001, 002 \& 003. The RCA separation between different ply orientations in Figure~\ref{fig:cmplx plots} is a result of fibres in plies at different depths contributing differently to the relative real and imaginary components in the ECT measurements. The RCA provides new information about the stacking sequence of the material as demonstrated by the changes in RCA of the same ply orientations across multiple samples (see Table~\ref{tab:pha}). In the following section~\ref{sec:struc diff 2}, this analysis technique is used to distinguish between samples A \& B from section~\ref{sec:AB}.

\subsection{Stacking-Sequence Differentiation: Complex}\label{sec:struc diff 2}

\begin{figure}[h]
\centering
    \includegraphics[width=0.75\textwidth]{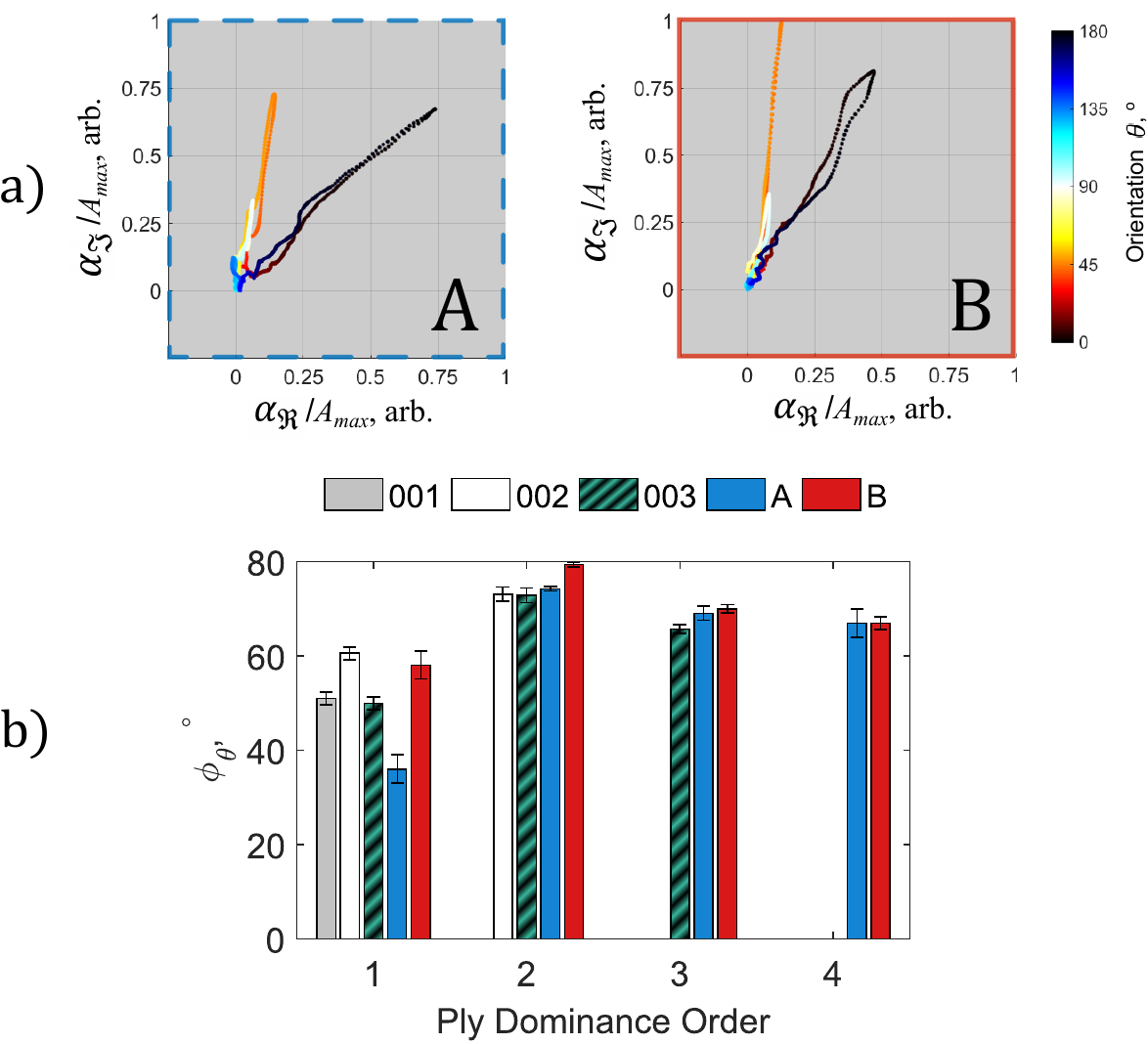}\caption{Real-imaginary RT analysis results for two CFRP samples: A - [$45^{\circ}$/$90^{\circ}$/$135^{\circ}$/$0^{\circ}$]$_{repeated}$ and B - [$0^{\circ}$/$45^{\circ}$/$90^{\circ}$/$135^{\circ}$/$0^{\circ}$]$_{reflected}$. Showing a) the full, normalised RT angular distributions in real-imaginary space, and b) the relative contribution angles (RCAs) of the dominant peaks in all test samples samples A \& B.}
    \label{fig:Phase Seq}
\end{figure}

Figure~\ref{fig:Phase Seq} compares the relative real-imaginary angular-distribution plots of samples A \& B exposing the differences between the two samples that were hidden in the magnitude-only analysis of section~\ref{sec:AB}. The results are displayed in two forms; the first (Figure~\ref{fig:Phase Seq}.a) shows the full normalised angular distribution of each sample in the real-imaginary plot, and the second figure (Figure~\ref{fig:Phase Seq}.b) compares the RCAs of the dominant peaks for the two samples, showing them in descending order of peak magnitude to demonstrate the correlation between ply depth, peak angular-distribution (dominance) and RCA.

%

Figure~\ref{fig:Phase Seq} shows that the greatest differences in RCA occur for the more dominant ply-orientations in spite of having the same ply orientations and very similar peak amplitudes (see Figure~\ref{fig:AngDistSamps}). It is the difference between these dominant orientations that allow the stacking sequences to be differentiated between. Samples A \& B exhibit strong differences in RCA for their two most dominant orientations ($0^{\circ}$ and $45^{\circ}$) of $21\pm2 ^{\circ}$ and $3\pm1^{\circ}$ respectively (see Figure~\ref{fig:Phase Seq} and Table~\ref{tab:pha}). This is an indication of the difference in the structure of samples A \& B. RCA analysis of complex ECT data could be used as a fingerprinting technique to determine the stacking sequence of CFRP structures. It is also important to note the trend in RCA as a function of ply dominance, where the most dominant peaks exhibiting the lowest RCAs with the second most dominant exhibiting the highest RCAs. This is likely linked to the fact that the current density is real at the surface of a test material and decays in amplitude and phase with depth (see equation~\ref{eqn:J}) such that the top layer will contribute more to the real component of the ECT measurement. Properly characterised, this analysis technique could be used to identify unknown layup structures and could be incorporated into in-process inspections to independently, and accurately check stacking sequence and alignment accuracy of multiple plies at a time. 

\begin{table}[H]
\caption[]{Features of real-imaginary angular-distribution plots. Averages from 36 separate analysis calculations made of 91$\times$91 pixel windows of each data set. Errors defined as 3 standard deviations about the mean.}
	\begin{center}\small	
		\begin{tabular*}{0.5\textwidth}{@{\extracolsep{\fill} } l c c c c}

			Sample & {} &  $\theta, \pm 0.1^{\circ} $ & $A_{\theta}$, arb. & $\phi_{\theta}, ^{\circ}$ \\
			\hline
			\hline
			001 & [90] & 89.6 & $2.2 \pm 0.1$ & $51 \pm 1$  \\
			\hline
			002 & [0] & -0.5 & $5.9 \pm 0.1$ & $73.1 \pm 0.5$  \\
			{} & [90] & 89.5 & $4.1 \pm 0.2$ & $61 \pm 1$  \\
			\hline
			003 & [45] & 44.7 & $2.3 \pm 0.1$ & $73 \pm 1$  \\
			{} & [90] & 89.6 & $4.0 \pm 0.2$ & $50 \pm 1$  \\
			{} & [135] & 133.7 & $2.1 \pm 0.1$ & $65.7 \pm 0.6$  \\
			\hline
			A & [0] & -0.4 & $6.0 \pm 0.2$ & $39 \pm 2$  \\
			{} & [45] & 45.5 & $4.6 \pm 0.1$ & $79 \pm 1$  \\
			{} & [90] & 90.5 & $3.1 \pm 0.1$ & $77 \pm 1$  \\
			{} & [135] & 134.2 & $1.22 \pm 0.02$ & $90 \pm 2$  \\
			\hline
			B & [0] & 0.6 & $7.4 \pm 0.3$ & $60 \pm 2$  \\
			{} & [45] & 44.5 & $5.8 \pm 0.2$ & $82 \pm 1$  \\
			{} & [90] & 90.3 & $3.1 \pm 0.1$ & $77 \pm 1$  \\
			{} & [135] & 136.9 & $0.8 \pm 0.03$ & $64 \pm 2$  \\
			\hline
		\end{tabular*}
		\label{tab:pha}
	\end{center}
\end{table}

\section{Conclusion}
High frequency eddy-current testing inspections of CRFP were used to produce high resolution images of the material structure. We demonstrated a Radon-transform analysis method applied to the complex inspection measurement data of an ECT scan to determine the dominant ply orientations and their distribution about this angle. The Radon-transform analysis method was shown to produce superior results compared to the commonly used 2D-fast Fourier transform analysis technique, and has the added advantage of eliminating a number of computational steps to quickly and quantitatively determine the ply orientations present in a structure. This analysis technique could be incorporated into automatic quality-control ECT inspections of pre-preg CFRP after lay-up to confirm the correct alignment and stacking sequence has been achieved. 

This study outlined a method for analysing the \textit{complex} behaviour of ECT inspection data using the Radon-transform. This technique conclusively demonstrated that more structural information is retrievable from full analysis of the complex data than merely the magnitude-only analysis methods commonly used. The method generated separate real and imaginary angular-distributions and plotted them on a relative real-imaginary angular-distribution plot, representing the relative changes in the original complex scan data. It was shown that different ply-orientations are distinguishable in the relative contribution angle in real-imaginary space, providing additional information about the sample-structure. Through the inspection and analysis of multiple CFRP structures, we have shown that this angular information is dependent on the stacking sequence of the CFRP structure, making it possible to distinguish between samples that are otherwise indistinguishable using magnitude-only analysis methods, but have different stacking sequences.

This work could lead to more accurate inspection techniques for composites at the pre-cure stage of the manufacturing process. Such a technique would allow re-work and guard against manufacturing errors that could result in unnecessary material losses or mechanical failure of critical components.

\section*{Acknowledgements}
The authors would like to thank Dr Luke Nelson and Ms Christina Fraij of the University of Bristol for their help with data analysis and sample manufacture respectively. This work is funded by the Engineering and Physical Sciences Research Council (EPSRC) [grant number EP/L015587/1]; and the Research Centre for Non-Destructive Evaluation [grant number EP/L022125/1]. The data created during this project can be found at the University of Bristol Research Data Storage Facility (RDSF) - DOI to be added after peer review.

\bibliography{RRHughes_Manuscript}




%

\end{document}